\documentclass[twocolumn,showpacs,preprintnumbers,amsmath,amssymb,floatfix,showkeys]{revtex4}

\usepackage{graphicx}
\usepackage{dcolumn}
\usepackage{bm}

\def\Includegraphics#1 {Here, include graphics #1}   

\def \D {\hbox{d}}
\def \barA {\overline{A}}

\def\geom{g} 

\def \de  {\triangle    \varepsilon}
\def \ee  {     \mathcal{E}}   
\def \intIm           {I_{\rm m}}
\def \intId           {I_{\rm d}}

\def \gnl{\gamma_{\rm NL}} 

\def \mod#1{\vert #1 \vert}
\def \exp {\mathop{\rm exp}\nolimits}

\def\vg{v_{\rm g}}
\def\vc{\overline{v}_{\rm g}}

\def\today{12~December~2011}
\def\today{20~December~2011}
\def\today{26~January~2012}
\def\today{3~May~2012 17:00}
\def\today{7~May~2012} 

\begin{document}

\preprint{CMLA--2012/xxxx}

\title{Nonlinear amplification of coherent waves \\ in media 
      with soliton-type refractive index pattern}     

\author{S.~Bugaychuk,$^{1,}$}
\email{bugaich@iop.kiev.ua}
\author{R.~Conte,$^{2,}$}
\email{Robert.Conte@cea.fr}

\affiliation{
$^1$Institute of Physics,
National Academy of Sciences,
46 Prospect Nauki, Kiev 03028, 
Ukraine,
}

\affiliation{
$^2$LRC MESO,\'Ecole normale sup\'erieure de Cachan (CMLA) et CEA--DAM
\\
61, avenue du Pr\'esident Wilson, F--94235 Cachan Cedex, France.
\\
Service de physique de l'\'etat condens\'e (CNRS URA 2464),
CEA-Saclay, F-91191 Gif-sur-Yvette Cedex, France
}
\date{\today}

\begin{abstract}
We derive the complex Ginzburg-Landau equation
for the dynamical self-diffraction of optical waves in a nonlinear cavity.
The case of the reflection geometry of wave interaction as well as a medium that 
possesses the cubic nonlinearity (including a local and a nonlocal nonlinear responses) 
and the relaxation is considered. 
A stable localized spatial structure in the form of a ``dark'' dissipative 
soliton is formed in the cavity in the steady state. 
The envelope of the intensity pattern, as well as of the dynamical grating amplitude, 
takes the shape of a $\tanh$ function.
The obtained complex Ginzburg-Landau equation 
describes the dynamics of this envelope, at the same time the evolution 
of this spatial structure changes the parameters of the output waves.
New effects are predicted in this system due to the transformation of the 
dissipative soliton 
which takes place during the interaction of a pulse with a continuous wave, such as:
retention 
of the pulse shape during the transmission of impulses in a long nonlinear cavity; 
giant amplification of a seed pulse, which takes energy due to redistribution of the pump 
continuous energy into the signal.

\end{abstract}


\maketitle

\section{Introduction}
\label{sec:Introduction}

Nonlinear systems which give different types of redistribution 
of matter or energy (intensity) are intensively studied in modern physics. 
Nonlinear processes can yield unusual forms of energy distribution. 
As a rule, they are described by nonlinear equations, 
such as the nonlinear Schr\"odinger equation (NLS), 
the complex Ginzburg-Landau equation (CGLE), their various modifications, 
or other \cite{AK2002,AABook2005,AABook2008,EPJ2010RW}.
It has been shown in many instances 
that the CGLE describes spatio-temporal localized structures, 
also called ``dissipative solitons'' 
\cite{AABook2005,AABook2008,PBA2010}.
The NLS is associated with a lot of physical phenomena, including rogue waves \cite{EPJ2010RW}.
In nonlinear systems amazing processes may exist, 
such as strong bursts of energy during the interaction of solitons \cite{MatveevRW2010}.

Mixing of several waves in a nonlinear medium very often leads to the redistribution of 
the wave intensities. 
As a rule, the parametric interaction processes are considered, when a monochromatic 
wave is amplified by a pumping wave which has another frequency \cite{Boyd,OPA2003}.
When the monochromatic waves interact in the nonlinear medium, 
one observes the effects of the self-action of the waves. 
The energy transfer is a peculiarity of the degenerate four-wave mixing (FWM) in media 
with a nonlocal response which takes place during self-diffraction of waves from 
the dynamical grating \cite{Staselko,OdoulBook,Landmark,Brignon}.
The background of this effect lies in a shift between the light interference pattern 
and a dynamical refractive index grating. 
This brings an additional phase shift between the transmitted and diffracted waves during their 
interference in the medium.
The intensity ratio of input waves is a control parameter,
which defines the magnitude of the energy transfer and amplification coefficient for a signal wave
\cite{HongSaxema,SBQuantum,Sturman2005}.

The FWM in media with a nonlocal response, 
when the self-diffraction of waves on a dynamical grating occurs, 
is one more physical system that is described by the CGLE \cite{BCPRE2009}.
In the present work we find the CGLE which describes the formation and evolution 
of the ``dark'' dissipative soliton along the longitudinal direction of wave propagation. 
The envelope of the refractive index distribution $\ee(t,z)$ has a soliton-like form.
All other characteristics of the FWM (wave intensities, diffraction efficiency) 
are expressed through this function $\ee(t,z)$.
The stable soliton-type pattern is obtained as a consequence of the nonlocal nature of the 
nonlinear response, 
which provides an additional phase shift for the waves diffracted by an 
inhomogeneous structure of the refractive index, 
and there is the interference between the diffracted and the propagating waves.							
The present research allows one to predict new effects of nonlinear wave interaction,
based on the properties of the soliton-like refractive index pattern.
As examples, we predict new effects, which take place during the interaction of a signal pulse 
and a continuous pump wave, and which are determined depending on suitable
initial conditions for the interacting waves and nonlinearity of the medium.
One of them is the retention of the shape of a signal pulse at the end of a bulk 
nonlinear medium, an application of which, for example, 
can be the transmission of pulses in long optical lines.
A second effect is a significant nonlinear amplification of a weak signal pulse at the expense 
of a continuous pumping wave, when a backward pumping wave is almost completely reflected from 
the nonuniform grating and redistributes its energy to a forward signal pulse. 
This phenomenon can be used for the creation of pulses of great energy.

\section{Derivation of the complex Ginzburg-Landau equation 
for the reflection four-wave mixing}  
\label{sec:II}

The four-wave mixing is a process which in the degenerate case
(all waves have the same frequencies) leads to many effects of self-action 
of waves during their interaction in a nonlinear medium.  
The phenomenon of the self-diffraction of waves during FWM combines three simultaneous processes: 
the recording of a 
time-dependent refractive index grating by the light interference pattern, 
the wave diffraction by the grating, 
and the interference between the propagated and diffracted waves. 
When the medium displays a nonlocal response due to some kind of transport mechanism,
one can write an evolution equation governing the dynamics of the induced nonlinearity, 
which is added to the FWM coupled wave system.

\begin{figure}
 \includegraphics{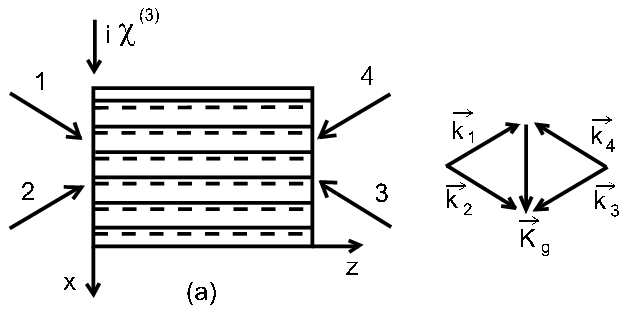} 
 \includegraphics{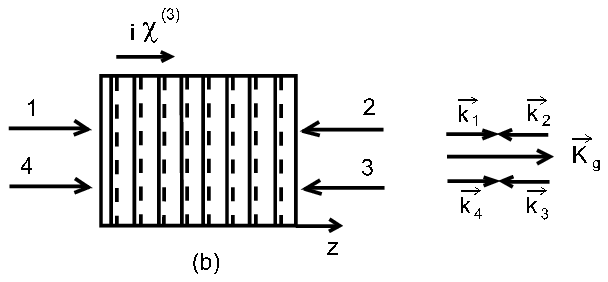}  
\caption{  \label{fig.refl.scheme} 
Scheme of degenerate FWM and $\vec{k}$ vectors diagrams. 
$\vec{K}_g=\vec{k}_1-\vec{k}_2=\vec{k}_4-\vec{k}_3$ is the grating vector.
(a) Transmission geometry. (b) Reflection geometry.
The numbers denote the input waves.
Full fines show the maxima of the interference pattern,
dashed lines show the maxima of the grating amplitude.
The arrow labeled $i \chi^{(3)}$ indicates the direction of 
a shift of the grating amplitudes relative to the maximum intensities of the light 
interference pattern.}
\end{figure}

The FWM has been previously investigated in the transmission geometry, 
when the coupled waves record the transmission dynamical grating and diffract on it 
(see FIG.~\ref{fig.refl.scheme}a).						
In nonlocal media both the intensity pattern and the grating amplitude distribution   
share the properties of a bright dissipative soliton 
\cite{BKMPR,BCPRE2009,JBHJOSAB,CBJPA2009,CBScicli}.
We have derived the cubic CGLE with a time dependent
gain/loss coefficient which describes the dynamics of the FWM in the restrictive case 
of a purely nonlocal response of the medium ($\gamma$ is a purely imaginary constant).
Meanwhile for the FWM in the reflection geometry (see FIG.~\ref{fig.refl.scheme}b)
the distribution of the intensity of the interference pattern shares the properties 
of a dark dissipative soliton. 

Transmission and reflection geometries differ in the direction of the 
grating-vector ($\vec{K}_g$) relative to the wave-vector components 
($\vec{k}_j$ , $j=1,2,3,4$). 
While in the transmission geometry the propagation of the waves is considered in the 
$(x,z)$ plane and the grating vector is directed just along the $x$-axis,
in the reflection geometry 
the problem is one-dimensional and all the vectors are parallel to the $z$-axis.
Also these two systems have different first integrals.
If one defines a symbol for describing the geometry
($\geom=1$ for transmission, $\geom=-1$ for reflection),
the FWM system in the nonlocal medium can be written in the unified way,
\begin{eqnarray}
& &
\partial_z     A_1=-i \ee     A_2,\
\partial_z \barA_2= \geom i \ee \barA_1,\
\nonumber\\ & &
\partial_z \barA_3=-i \ee \barA_4,\
\partial_z     A_4= \geom i \ee     A_3,\
\label{eqFWM}
\\ & &
\partial_t \ee = \gamma \frac{\intIm}{I_0} - \frac{\ee}{\tau},
\label{eqgrating_t}
\\ & &
\intIm =A_1 \barA_2 + \barA_3 A_4,
\label{eqdefIm}
\\ & &
I_0 = \mod{A_1}^2 + \mod{A_2}^2 + \mod{A_3}^2 + \mod{A_4}^2, 
\label{eqdefI0}
\\ & &
\intId= -\mod{A_1}^2 + \mod{A_2}^2 + \geom (- \mod{A_3}^2 + \mod{A_4}^2),
\label{eqdefId}
\end{eqnarray}
where $A_j, j=1,2,3,4$ are the slow variable amplitudes of interacting waves, 
$\ee$ is the amplitude of the grating, 
$\gamma$ is a complex constant which describes the maximum amplification of the medium, 
$\tau$ is a time relaxation constant of the grating,
$\intIm$ is the interference pattern, 
$I_0$ is the total intensity, 
$\intId$ is the relative net gain.
The first integral is $I_0$ in the transmission geometry
and $\intId$ in the reflection geometry.
Equation (\ref{eqgrating_t}) is the evolution equation, 
where for simplicity we include only two terms \cite{OdoulBook}: 
amplification of the grating amplitude proportional to the light intensity at every local point $z$,
and usual exponential (dielectric) relaxation of the dynamical grating. 

The coupled wave eqs.~(\ref{eqFWM}) for slow variable amplitudes are derived 			
from the Maxwell's wave equation, 
taking into consideration that the amplitude of the dynamical grating is determined as 
$\ee=\de \exp(i \vec{K}_g \vec{r})$ + c.c. 
In the approximation of small variations of the refractive index we have
$\de \cong 2 n_0 \Delta n$ , 
where $\de$ and $\Delta n$ are respectively the variations of 
the dielectric permittivity and the refractive index of the medium induced by the laser radiation, 
and $n_0$ is the average refractive index in the medium. 
In our consideration we take into account that the nonlinear gain coefficient 
is a complex constant $\gamma = \gamma_L + i \gnl $,
where $\gamma_L$  and $\gnl$ describe, respectively, the local and 
nonlocal responses of the medium.
The gain coefficient in a nonlocal medium can be written as \cite{OdoulBook}
$\gamma = 2 \pi \Delta n_{\rm max} (\cos \Phi_g + i \sin \Phi_g)/\lambda$, 
where $\Phi_g$ describes a shift of the dynamical grating with respect 
to the maxima of the interference pattern (minimal spacing between maxima 
of the light lattice and the refractive index grating in a positive direction 
of a polar (unidirectional) axis of the medium), and $\Delta n_{\rm max}$  
is the maximally possible grating amplitude in the given medium.
Both these values are determined by the physical mechanisms that take place 
during the grating recording.
The complex values of $\gamma$, $\de$  and $\ee$ show that 
a diffracted wave gets some additional phase shift relative to a propagated 
wave after the diffraction from the refractive index grating.
In the case of a purely nonlocal response ($\gamma_L=0$ , $\gamma = i \gnl$) 
this phase shift equals $-\pi /2$ for the wave diffracted in the direction 
of the polar axis and $\pi/2$  for the wave diffracted in the opposite direction.
In this special case in the steady state, the wave equations (\ref{eqFWM}) 
becomes real, where the waves 1 and 4 are amplified at the expense of the waves 2 and 3, 
which transfer their energy to the waves 1 and 4 respectively.
The energy transfer between interacting waves takes place if the medium has a purely
nonlocal response.
When the response is complex and includes a local component, 
both a phase transfer and an energy transfer occur
between the interacting waves.

We assume the following normalization in the system (\ref{eqFWM})--(\ref{eqdefId}):
the variable $\ee$ is dimensionless, 
the coefficient $\gamma$ has the dimension $[\gamma]=T^{-1}$, 
and the independent variable $z$ is normalized as
$z=[k_0^2/(2k_{z\acute{}})]z\acute{}$, where 
$k_0$ is the amplitude of the wave-vector in the free space, 
$z\acute{}$ is the longitudinal spatial coordinate and 
$k_{z\acute{}}$ is $z\acute{}$-component of the wave-vector in the nonlinear medium.

Like in the transmission geometry \cite{BCPRE2009}, 
the system (\ref{eqFWM})--(\ref{eqdefId}) in five complex variables 
in the reflection geometry
is reducible to an intrinsic system in two complex variables
$\ee$ and $J_m=\intIm /I_0$,
\begin{eqnarray}
& &
\partial_t \ee = \gamma J_m - \ee / \tau  ,\
\label{eqIntr1}   
\\ & &
\partial_z J_m = -i \ee - 2i \bar{\ee} J_m J_m + 2i \ee J_m \bar{J_m}.
\label{eqIntr2}
\end{eqnarray}
Eliminating $J_m$ between eqs.~(\ref{eqIntr1})--(\ref{eqIntr2}) yields
\begin{eqnarray}
E \equiv 
\partial_t \partial_z \ee + \frac{1}{\tau} \partial_z \ee + i \gamma \ee 
 - \frac{2 i}{\tau \left|\gamma \right| ^2} \times \
\nonumber \\   
\left[ \frac{\bar{\gamma} - \gamma}{\tau} \left|\ee \right|^2 \
+ \gamma \ee \partial_t \bar{\ee} - \bar{\gamma} \bar{\ee} \partial_t \ee \right] 
   \times \left[\ee + \tau \partial_t \ee  \right] = 0.   
\label{eqLast}
\end{eqnarray}
This is our final equation, to which we apply the reductive 
perturbation method to obtain the CGLE.
Like in \cite{BCPRE2009, DauxoisPeyrard}, 
we define a multiple scale expansion
in which the function $\ee$ is of order $\varepsilon$,
where the function $\varphi_k$ depends on a set of variables associated with 
these various scales:
\begin{eqnarray}
& &
\ee(z,t) = \varepsilon \sum_{k=0}^{+\infty}
\varepsilon^k 
\varphi_k(Z_0,\dots,Z_k,\dots,T_0,\dots,T_k,\dots),
\nonumber\\ & &
Z_k=\varepsilon^k z,\ T_k = \varepsilon^k t,\
E=\varepsilon \sum_{k=0}^{+\infty}
\varepsilon^k E_k,        
\label{eqphi}
\end{eqnarray}   
and we require each coefficient $E_j$ to vanish.
														
The method of a multiple scale expansion applies to weakly dispersive and weakly nonlinear 
systems which are described by a wave equation in the small-amplitude limit.
It was used to derive the NLS equation, which shows behavior of envelope solitons.  
This method permits one to impose appropriate conditions in the multidimensional space 
that eliminate the divergences of the asymptotic expansion for small values of $\varepsilon$
(see chapter 3 in \cite{DauxoisPeyrard}).


At zero-th order, the equation for the function $\varphi_0$:
\begin{eqnarray}
& &
L \varphi_0=0,\
L \equiv
\partial_{T_0} \partial_{Z_0} + \frac{1}{\tau}\partial_{Z_0} + i \gamma,
\label{eqOrder0}
\end{eqnarray}
admits for solution the complex plane wave,
\begin{eqnarray}
& & {\hskip -11.0truemm}
\varphi_0=A(Z_1,Z_2,T_1,T_2,\dots) e^{\Phi_0},\ 
\Phi_0=i(q Z_0-\omega T_0), 
\label{eqOrder0OneWave}
\end{eqnarray} 
where the phase $\Phi_0$ depends on the variables $T_0$ and $Z_0$,
the amplitude factor $A$ is a function depending on the other space and time scales, and
the constants $q, \omega$ obey the dispersion relation
\begin{eqnarray}
& &
q \omega + i \left( \frac{q}{\tau} + \gamma \right) =0.
\label{eqdisper}
\end{eqnarray}
{}From this equation it follows that the wave-vector $q$ and the pulsation $\omega$ 
take complex values, a consequence of both the relaxation $1/\tau$ and the complex 
nature of $\gamma$.

At first order, in the equation for $\varphi_1$
\begin{eqnarray}
& & {\hskip -8.0truemm}
L \varphi_1=-       G_{1}  e^{          \Phi_0}
        - \overline{G_{1}} e^{\overline{\Phi_0}},\
\nonumber
\\
& &   
G_{1} \equiv
i q               \frac{\partial A}{\partial_{T_1}}
- \frac{\gamma}{q} \frac{\partial A}{\partial_{Z_1}},
\end{eqnarray}
$G_1$ must vanish to avoid $\varphi_1$ to diverge,
providing 
\begin{eqnarray}
& & 
\varphi_1=0,\
A= \Psi (Z_1-\vg T_1,Z_2-\vg T_2,T_2,\dots),\ 
\label{eqEps1}
\end{eqnarray}
in which the
group velocity $\vg= i \gamma / q^2$ is generically complex
and
the complex function of integration $\Psi$ is 
to be determined.
Let us introduce the two complex conjugate independent variables $X_1,Y_1$,
\begin{eqnarray}
& &
X_1=Z_1-\vg T_1,\
Y_1=\overline{X}_1=Z_1-\vc T_1.
\end{eqnarray}

In the second order equation for the evolution of $\varphi_2$,
\begin{eqnarray}
& &
L \varphi_2=-       G_{2}  e^{          \Phi_0} 
            - \overline{G_{2}} e^{\overline{\Phi_0}},\
\end{eqnarray}
the cancellation of the secular terms requires $G_2$ to vanish,
which defines two complex conjugate nonlinear PDEs for 
$\Psi(X_1,T_2)$ and $\bar \Psi(Y_1,T_2)$, 
and yields the value $\varphi_2 =0$.
The resulting equation $G_2=0$ is the desired CGLE,
\begin{eqnarray} {\hskip -2.0truemm}     
G_{2} \equiv
  i \frac{\partial \Psi}{\partial T_2} 
  - i \frac{\gamma}{q^3} \frac{\partial^2 \Psi}{\partial X_1^2}
  - 4 \frac{\gamma}{q^3} \frac{\Im(q)}{\bar{q}} e^{2\Re (\Phi_0)} \left|\Psi \right|^2 \Psi =0, 
\label{eqCGLERef}
\end{eqnarray}
where 
$\Re (\Phi_0)$ denotes the real part of the phase $\Phi_0$,
$\Im(q)$ indicates the imaginary part of the wave vector $q$.
The obtained CGLE describes the spatio-temporal dynamics 
of the grating amplitude envelope during the FWM
in a nonlinear medium with a complex response (both local and nonlocal). 
The complex coefficients in the above CGLE arise from the relaxation 
of the photoinduced refractive index
as well as from the local component of the nonlinear response.
The same dynamics will be for the distribution of the intensity in the medium.
A variety of solutions of the CGLE (\ref{eqCGLERef}) 
will give possible localized structures and their behaviors 
that can be implemented during the reflection FWM.

\section{Alteration of the grating amplitude distribution in the steady state}
\label{section:III}

\begin{figure}
 \includegraphics{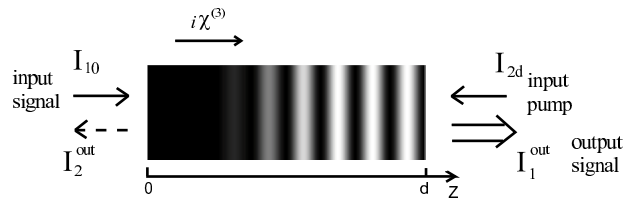} 
\caption{  \label{fig.refl.localized_structure} 
Localized structure of the dark dissipative soliton.} 
\end{figure}

In the simplest case, 
the four-wave mixing in the reflection geometry can be reduced to the two-wave mixing (TWM), 
which describes the interaction of forward and backward waves. 
Then the FWM momentum conservation law takes the form  $2k_{1} = 2k_{2}$, where $k_{1}$, 
$k_{2}$  are the $z$-components of the wave vectors for the forward and backward waves 
respectively.
In the next sections we consider the effects which arise in this simplest case of 
the TWM in a medium with a purely nonlocal response $\gamma = i \gnl$, 
i.e.~there is a 
constant space shift,
equal to one quarter of the period of the light interference pattern, 
between the fringe interference pattern and the photoinduced dynamical grating.
We consider the case where the grating is shifted in the $+z$  direction,
so the signal beam (entered on the input boundary $z=0$) is amplified, 
but the backward beam (entered on the boundary $z=d$) is the pump beam. 
In the case of a purely nonlocal response,
the system (\ref{eqFWM})--(\ref{eqdefIm}) simplifies to a set of real equations.
In the steady state the distribution of the grating amplitude is then  
connected 
with the maxima of the interference pattern by the following relation,
\begin{eqnarray}
\ee(z)=\gnl \tau \intIm (z)/I_{0}(z).
\label{eq3_1}
\end{eqnarray} 
The solution for the grating amplitude is
\begin{eqnarray}
\ee(z)= \frac{1}{\sqrt{2}} \sqrt{1 + \tanh \left(\gnl \tau z 
 + \frac{1}{2} \log \left( \frac{4}{\intId^2} \right) -p \right)},
\label{eq3_2}
\end{eqnarray} 
where the first integral is $\intId^{2} = I_{0}^{2} - 4 \intIm^{2}$, 
and the integration constant $p$ can be found from the input boundary conditions.
If one takes account of (\ref{eq3_1}), 
the normalized intensity pattern $\intIm/I_{0}$  
has the same distribution (\ref{eq3_2}) (up to the constant $\gnl \tau $), 
but the patterns $\intIm/I_{0}$ and $\ee$ are shifted by a phase of $\pi /2$ 
(equivalent to the space shift of $\Lambda /4$, 
where $\Lambda$ is the period of the interference pattern along the longitudinal $z$-direction). 
The spatial structure of the nonuniform intensity pattern is shown 
in FIG.~\ref{fig.refl.localized_structure},
where the inflection point of the $\tanh$ function is located inside the bulk of the nonlinear medium.

By changing the input intensity ratio, 
the envelope $\intIm(z)/I_0(z)$ (as well as $\ee(z)$) 
moves along the $z$-direction without changing its form (see FIG.~\ref{fig.refl.localized_structure}).
So, one can define this spatial nonuniform pattern as the dark dissipative soliton.
Modulation of the refractive index takes the same structure as the dark dissipative soliton, 
in which the interacting waves undergo diffraction. 
The stationary solutions for the wave amplitudes are
\begin{eqnarray}
& &
A_1(z)=C_1 e^{U(z)} + C_2 e^{-U(z)},\
\nonumber\\ & &
A_2(z)=C_1 e^{U(z)} - C_2 e^{-U(z)}.
\label{eq3_3}
\end{eqnarray}
They are determined by the area $U(z)$
under the curve of the grating amplitude envelope,
\begin{eqnarray}
& &
{\hskip -10.0 truemm}
U(z)= \int\limits_{0}^{z} \ee(z) \D z 
=\frac{1}{4} \ln \left( \sqrt{\left(e^w \right) ^2 + e^w } + e^w + \frac{1}{2} \right),
\label{eq3_4}
\end{eqnarray}
where $w=2 \gnl \tau z + \log \left(4/\intId^2 \right)-2p$. 
Denoting $d$ the thickness of the medium,
$A_{10}=A_1(z=0)$, $A_{2d}=A_2(z=d)$ the amplitudes of the  input waves,
$A_1(d)=A_1(z=d)$, $A_2(0)=A_2(z=0)$ the amplitudes of the output waves, 
and 
$U_d$ the area under the grating amplitude envelope within the whole medium 
$U_d=\int^{d}_{0} \ee(z) \D z $,
the constants of integration in (\ref{eq3_3}) evaluate to 
\begin{eqnarray}
& & {\hskip -9.0truemm}
C_1=\left(A_{10} e^{-U_d}+A_{2d}\right)/(2 \cosh U_d),\
\nonumber\\ & & {\hskip -9.0truemm}
C_2=\left(A_{10} e^{ U_d}-A_{2d}\right)/(2 \cosh U_d),\
\intId =4 C_1 C_2,
\label{eq3_5}
\end{eqnarray}
and the constant $p$ is determined by the conditions at the medium boundary,
\begin{eqnarray}
& &
\ee(0)=\gnl \tau \frac{A_{10} A_2(0)}{A_{10}^2+A_2(0)^2} 
\\ \nonumber & &
=\frac{1}{2}
\sqrt{1+\tanh \left(\frac{1}{2} \log \left( \frac{4}{\intId^2} \right) -p \right)}.
\label{eq3_6}
\end{eqnarray}

\begin{figure}
 \includegraphics{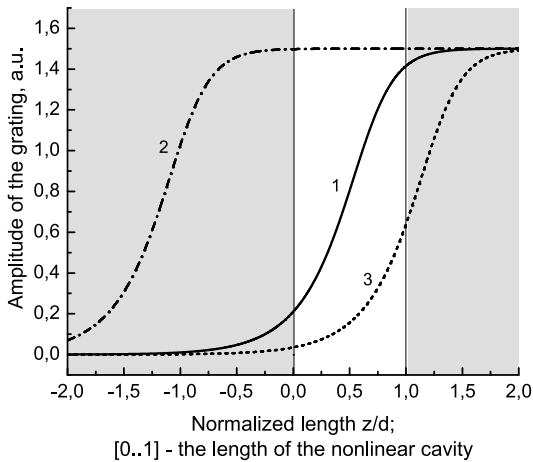} 
\caption{  \label{fig.refl.grating_vs_I_ratio}                            
Stationary distribution of the grating amplitude for different
input intensity ratios. 
1: $I_{10}/I_{2d}=1,    K_{\rm sig}= 2$; 
2: $I_{10}/I_{2d}=0.05, K_{\rm sig}=20$; 
3: $I_{10}/I_{2d}=1.2,  K_{\rm sig}= 1$.
The normalized coupling constant of the medium is $\Gamma = \gnl \tau d =3$.} 
\end{figure}

{}From the solution (\ref{eq3_3}),
one sees that the intensity of the output signal wave $I_1^{\rm out}$
is determined by the area under the grating amplitude envelope 
between the boundaries of the nonlinear medium.
The spatial location of the grating envelope depends on the input intensity ratio 
$I_{10}/I_{2d}$, as seen from FIG.~\ref{fig.refl.grating_vs_I_ratio}.

\vfill\eject\clearpage 

The inflection point of the $\tanh$ function of $\ee(z)$ is located outside the boundaries 
for small signal and big pump intensity.
The grating amplitude has uniform distribution within the boundaries, 
and the gain coefficient of the signal ($K_{\rm sig}=I_1^{\rm out}/I_{10}$) is 
maximal possible in a given nonlinear medium.
When one increases the signal beam intensity as compared to the pump beam intensity, 
which is accompanied by a motion of the grating amplitude envelope, 
the inflection point of the $\ee(z)$ $\tanh$ function is already located inside the boundaries,
so that the nonuniform distribution of the grating amplitude is formed inside the 
nonlinear medium with the maximum being located close to the output boundary $z=d$.
The area under the envelope will decrease, then the amplification coefficient
of the signal beam decreases too. 
In this way we show that, by changing the input intensity ratio, 
the coefficient of the energy transfer will also be changed.
The reason for that is the motion of the grating amplitude envelope and the self-formation 
of either a uniform distribution or a nonuniform structure of the grating amplitude in the 
nonlinear medium.
At the same time the most pronounced effects connected with the alteration of the signal 
beam will occur when the nonuniform dynamical grating is located inside a nonlinear medium.      
Then small changes of the intensity ratio will lead to the motion of the grating envelope 
and therefore to significant changes of the output signal.

\begin{figure}
 \includegraphics{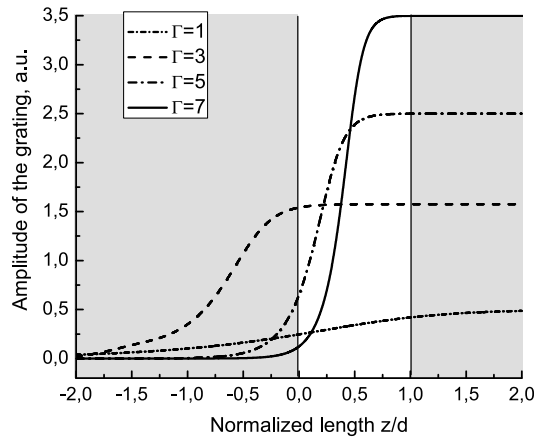} 
\caption{  \label{fig.refl.grating_vs_Gamma} 
Stationary distribution of the grating amplitude for
different normalized coupling constants $\Gamma = \gnl \tau d$.
$I_{10}/I_{2d}=0.05$.
}
\end{figure}

The amplification coefficient is also determined by the coupling constant, 
which complies with the maximum value of the nonlinearity in the optical cavity.
In the steady state the coupling constant is $\Gamma = \gnl \tau d$.
FIG.~\ref{fig.refl.grating_vs_Gamma} displays the dependence of the grating amplitude envelopes 
on the value of $\Gamma$.
This figure displays the following remarkable feature peculiar to just the reflection 
geometry of the wave-mixing:
the nonunifrom distribution of the grating envelope 
may be created in media with small or average nonlinear coefficients. 
Therefore, in order to implement the effects of beam control,
large nonlinearities of the medium are unnecessary.
Such effects are more efficient in media with small nonlinearities.

\section{Pulse propagation in the TWM}
\subsection{Transmission of a pulse retaining the pulse shape}
\label{section:IV.1}
																							
The motion of the soliton-like envelope of the dynamical grating owing to changes 
of the input intensity ratio becomes a very fruitful feature to archive various kinds 
of manipulations of laser pulses.
Here we consider two effects which take place during the interaction 
of a pumped continuous wave with a pulse signal in a nonlinear cavity, 
where a phase-shifted dynamical holographic grating is created in some way. 
We choose a Gaussian beam with respect to time in the form
$I_{10}(t)=I_{\rm b}+I_{\rm sig} \exp \left(-t^2/\tau_{\rm sig}^2 \right)$, 
where
$I_{\rm b}$ is the intensity of the background, 
$I_{\rm sig}$ is the maximum of the signal pulse and 
$\tau_{\rm sig}$ is its duration.
In this way, we consider the case of the self-diffraction of a signal pulse on a given grating, 
which is created by a small background intensity $I_{\rm b}$ 
of the wave 1 and by the continuous pump wave 2.
We point out that the background can be negligibly small 
(e.g.~a scattered wave, a reflected wave), 
but it should exist to build up the given grating.
We will show that the output signal will depend 
on the intensity ratio between the pump and the maximum signal, 
on the coupling coefficient of the  nonlinear medium as well as 
on the properties of the signal pulse itself, i.e.~on its duration and even on its shape. 

\begin{figure}
 \includegraphics{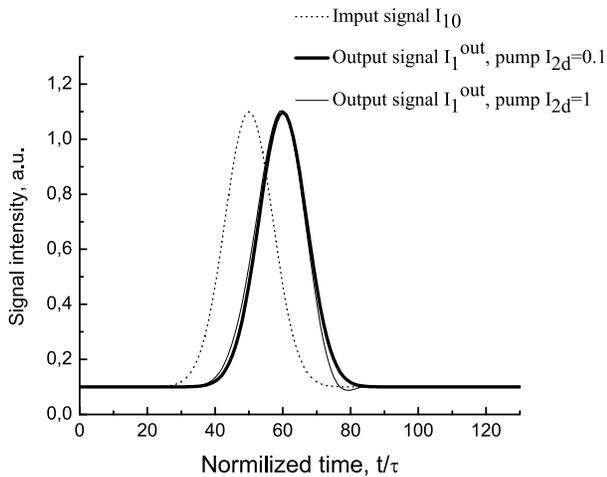} 
\caption{  \label{fig.refl.retention}  
Retention of the shape of the Gaussian pulse during its propagation in 
any long optical cavity, which contains the refractive index grating
having the ``dark'' soliton-like structure.  
The parameters of the Gaussian signal are 
$I_{\rm sig}=1$, 
$I_{\rm b}=0.1$, 
$\tau_{\rm sig}/\tau=10$ 
(all intensities are given in arbitrary units),
the normalized coupling constant is $\Gamma = \gnl \tau d=5$. 
}
\end{figure}

\begin{figure}
 \includegraphics{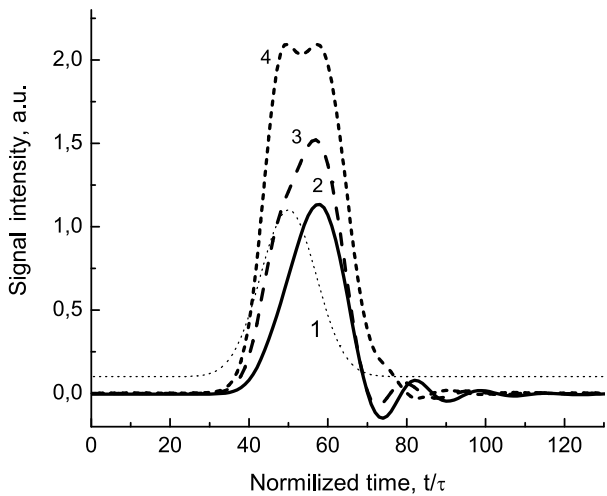}  
\caption{  \label{fig.refl.amplification} 				
Amplification and distortion of the Gaussian pulse during TWM upon increase of the pump intensity. 
1 - the intensity of the input signal, 
2 - the output signal $I_1^{\rm out}$ for the pump intensity $I_{2d}=10$, 
3 -                   $I_1^{\rm out}$ for                    $I_{2d}=30$, 
4 -                   $I_1^{\rm out}$ for                    $I_{2d}=50$. 
The parameters of the signal pulse and the coupling constant are identical to those 
in FIG.~\ref{fig.refl.retention}. 
}
\end{figure}

In the case of a weak pump the signal pulse retains its shape on the output of the cavity.
In FIG.~\ref{fig.refl.retention} we show that the shape of the output signal 
coincides with the shape of the input Gaussian signal 
when the pump intensity is smaller than or comparable to the maximum signal intensity 
$\left( I_{2d}\leq I_{\rm sig} \right)$.
One can transfer pulses in a long transmission line without distorsion of the pulse shape,
provided one creates the conditions that a weak backward scattering wave is created 
and involved in the recording of the shifted dynamical grating inside this transmission line.
When the pump is increased, 
the output signal is amplified
and the pulse shape is distorted (see FIG.~\ref{fig.refl.amplification}). 

\begin{figure}
 \includegraphics{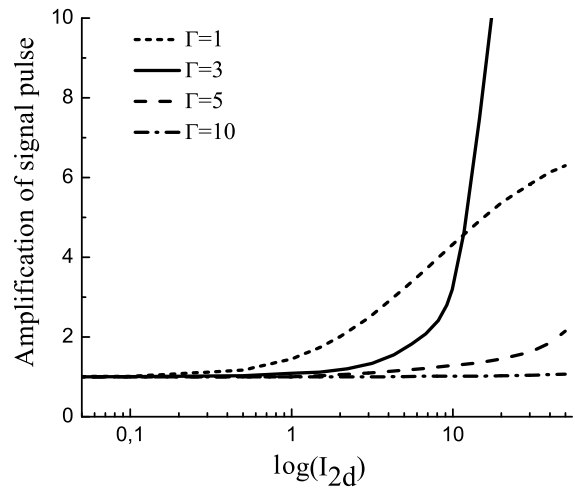} 
\caption{  \label{fig.refl.gain_vs_Gamma}                                  
Dependence of the gain of the signal pulse intensity on a small pumping 
for different coupling constants. 
The parameters of the signal pulse are identical to those in FIG.~\ref{fig.refl.retention}. 
}
\end{figure}

The effect of the signal amplification depends strongly 
on the value of the photoinduced optical nonlinearity, 
but in a nonobvious way: 
the amplification can be lowered by increasing the coupling constant $\Gamma$ 
(see FIG.~\ref{fig.refl.gain_vs_Gamma}). 
At the same time there exists an optimal value of $\Gamma$ 
when the maximum amplification is reached.
The explanation of this phenomenon is the same as that discussed 
in Section \ref{section:III} for the TWM in the steady state: 
for high values of $\Gamma$ the grating amplitude distribution is uniform 
over a volume of the nonlinear medium.
In this case, changes of the input intensity ratio have little influence
on the redistribution of the grating amplitude, 
this will have little effect on the intensity of the output signal.
Big changes can be reached when the distribution of the grating amplitude 
is not uniform over the volume of the medium.
In such a case small changes of the input intensity ratio 
lead to significant redistributions of the grating amplitude, 
and as a result the output signal will undergo large changes as well. 

\begin{figure}
 \includegraphics{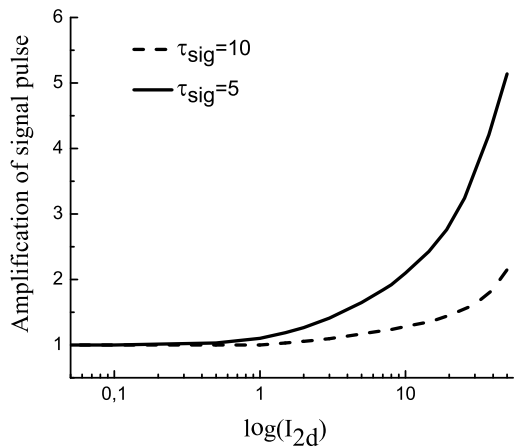} 
\caption{  \label{fig.refl.gain_vs_half-width}                                     
Dependence of the gain on the half-width of the Gaussian pulse,
for the coupling constant $\Gamma =5$.
}
\end{figure}

The gain for the reflection TWM depends on the steepness 
of the fronts of the signal beam. 
In the case of a Gaussian signal beam, 
the gain coefficient increases when the half-width of the pulse decreases
(see FIG.~\ref{fig.refl.gain_vs_half-width}).
The shape of the output pulse will also change significantly.
Features connected with the amplification of short laser pulses 
in the reflection TWM scheme are considered in the next subsection.

\subsection{Giant amplification of a short laser pulse}
\label{section:IV.2}											

\begin{figure}[h]
 \includegraphics{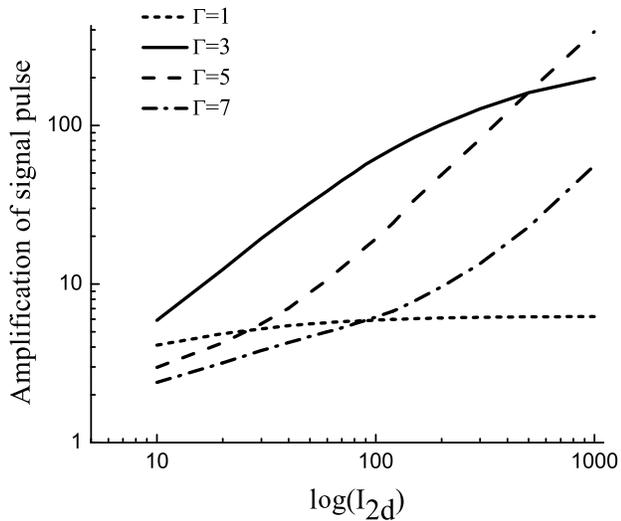} 
\caption{  \label{fig.refl.gain_Gaussian_vs_Gamma}
The gain of a short Gaussian pulse for different values of the coupling constant. 
$\tau_{\rm sig}/\tau =3$,  $I_{\rm sig}=1$, $I_{\rm b}=0.1$.
}
\end{figure}

Amplification of a weak short pulse depending on the pump intensity 
and the maximum value of the nonlinearity 
in the case of large pumping is shown in FIG.~\ref{fig.refl.gain_Gaussian_vs_Gamma}. 

\begin{figure}
 \includegraphics{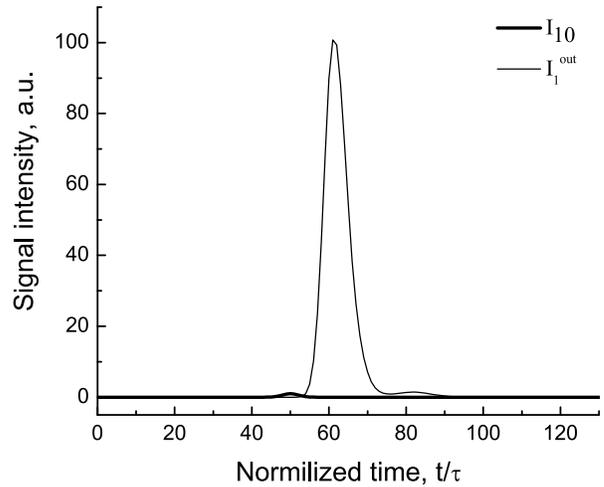}     
\caption{  \label{fig.refl.giant} 
Giant amplification of a short seed pulse for optimal values of the coupling constant. 
The parameters of the input signal pulse are identical to those 
in FIG.~\ref{fig.refl.gain_Gaussian_vs_Gamma};  
the pump intensity is $I_{2d}=200$ and the coupling constant is $\Gamma =3$.
}
\end{figure}

Like in FIG.~\ref{fig.refl.gain_vs_Gamma} this dependence is not one-to-one 
with respect to the coupling constant. 
The amplification of the signal can be small for both low and high values of $\Gamma$.
The gain coefficient becomes optimal for certain small values of $\Gamma$.
In this specific case the output signal may almost reach the value of the pump intensity. 
In FIG.~\ref{fig.refl.giant} we show the giant amplification of a seed Gaussian pulse 
due to high pumping in the range of the optimal values of the coupling constant.

\vfill\eject\clearpage 
\begin{figure}
 \includegraphics{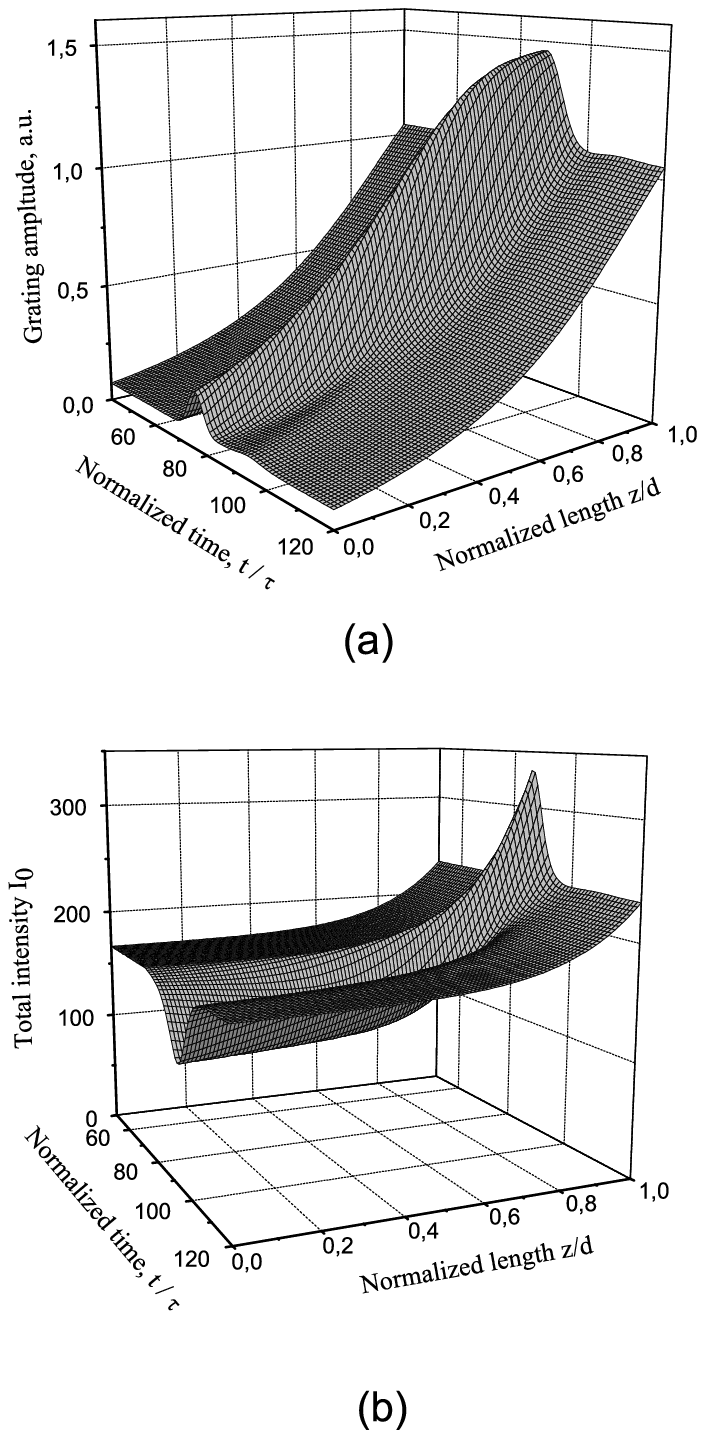}  
\caption{  \label{fig.refl.changes} 			
Changes of the grating amplitude envelope 
(a) 
and redistribution of the total intensity inside the cavity 
(b) under the TWM of a seed pulse with strong pumping in a bulk nonlocal medium.
The parameters used in calculation are identical to those in FIG.~\ref{fig.refl.giant}.
}
\end{figure}
In FIG.~\ref{fig.refl.changes} we calculate the rebuilding of the grating amplitude and the redistribution 
of the total intensity $I_0$ during the propagation of the pulse in the cavity, 
where the nonlocal dynamical grating is created. 
As one can see, they are both redistributed compared to their background values 
when a weak pulse appears on the input boundary.
In this case, the total intensity is redistributed inside the cavity in such a way that 
its very big maximum is concentrated close to the output boundary $z=d$,
whereas its very pronounced minimum is located near the input boundary $z=0$. 
In other words, this effect can be imagined as:
all the pump wave is reflected from the dynamical grating 
being photoinduced owing to a weak signal beam, 
and this is displayed as an expanded and significantly enhanced output pulse. 

\section{Conclusion}
\label{section_conclusion}

We have developed a model of formation of dissipative spatial soliton  
which takes place during the interaction and self-diffraction of coherent waves in media 
with a nonlocal nonlinear response.
We consider the media with cubic nonlinearity, 
and only two effects are taken into account, namely the light-induced modulation 
of the refractive index being proportional to the light intensity 
and the temporal relaxation of the refractive index dynamical grating.
In the simplest case of the fringe interference pattern, 
and when a nonlocal dynamical grating is shifted in space relatively to the light pattern, 
the envelope of the maximum amplitudes takes a soliton-like form created
along the direction of wave propagation $z$.
The spatial structure of the interference pattern ($\intIm(z)/I_0(z)$)
has the form of a dark dissipative soliton in the case of 
reflection geometry of wave interaction.
The same spatial pattern occurs for the distribution of the amplitudes of the grating
$\ee(z)$.
We have derived the complex Ginzburg-Landau equation, which describes the dynamics of  
self-formation of stable dissipative soliton as well as its evolution 
when the input conditions are changed. 
The expansion of this model for the case of interaction and diffraction 
of noncoherent waves with different frequencies is of undoubted interest.

We have explained that the coefficient of energy transfer depends on 
whether a uniform or a nonuniform distribution of the grating amplitude is formed
within the volume of the nonlinear medium.
This, in turn, is determined by the intensity ratio of the input waves.
We have found two interesting effects arising because of the interaction of a signal 
pulse with a continuous pump illumination in a nonlocal medium, 
the reason for that being a redistribution of the dynamical grating.
The first effect is the restoration of the form of the input pulse on the output                              
which takes place when the pump intensity is either comparable to or less than the 
maximum of the signal pulse.
But when one increases the pump intensity as compared to the signal, 
one observes a significant nonlinear amplification of a seed pulse.
This effect is due to the fact that the seed pulse provokes the creation of 
such a grating, 
which reflects almost the entire pump wave in the direction of the signal.
Different types of applications can be used in the system depending on suitable input conditions, 
for example, transmission of pulses over long distance in fiber amplifiers, 
or significant amplification of short pulses in a nonlinear optical cavity.

\begin{acknowledgments}

We gratefully acknowledge the financial support of the
Max-Planck-Institut f\"ur Physik komplexer Systeme,
where most of this work was performed.

\end{acknowledgments}


\begin{thebibliography}{99}

\bibitem{AABook2005} N.~Akhmediev and A.~Ankiewicz (eds.),
\textit{Dissipative solitons},
448 pages, Lecture notes in physics {\bf 661} (2005).

\bibitem{AABook2008} N.~Akhmediev and A.~Ankiewicz (eds.),
\textit{Dissipative solitons: from optics to biology and medicine},
Lecture notes in physics {\bf 751} (2008). 

\bibitem{EPJ2010RW} N.~Akhmediev and E.~Pelinovsky (eds.),
\hfill\break\noindent
Eur.~Phys.~J.~Special Topics {\bf 185} (2010).

\bibitem{AK2002} I.S.~Aranson and L.~Kramer,                 
Rev.~Mod.~Phys.~{\bf 74}, 99 
(2002).

\bibitem{PBA2010} H.-G.~Purwins, H.U.~B\"odeker and Sh.~Amiranashvili, 
\textit{Dissipative solitons}, 
Advances in Physics {\bf 59} 485 
(2010). 

\bibitem{MatveevRW2010} P.~Dubard, P.~Gaillard, C.~Klein and V.B.~Matveev,
Eur.~Phys.~J.~Special topics {\bf 185} 247 
(2010). 

\bibitem{Boyd} R.W.~Boyd, 
\textit{Nonlinear optics} (Academic Press, Boston, 2003).

\bibitem{OPA2003} G.~Cerullo and S.~De Silvestri,
Rev.~Sci.~Instrum.~{\bf 74} 1 
(2003).

\bibitem{Staselko} D.I.~Stasel'ko and V.G.~Sidorovich,  
\hfill\break\noindent 
Sov.~Phys.~Tech.~Phys.~{\bf 19} 361 
(1974) 
\hfill\break\noindent 
       [Zh.~Teck.~Fiz.~{\bf 44} 580 
       (1974)].
       
\bibitem{OdoulBook} S.~Odoulov, M.~Soskin and A.~Khizhnyak,
\textit{Optical oscillators with degenerate four-wave mixing},
(Harwood, Chur, Switzerland, 1991).  
       
\bibitem{Landmark} Pochi Yeh and Claire Gu (eds.),
\textit{Landmark papers on photorefractive nonlinear optics} 
(World Scientific, Singapore, 1995).     
       
\bibitem{Brignon} A.~Brignon and J.-P.~Huignard (eds.),
\textit{Phase conjugate laser optics} (Wiley, New York, 2004).

\bibitem{HongSaxema} J.H.~Hong and R.~Saxema,
Opt.~Lett.~{\bf 16} 180 
(1991).

\bibitem{SBQuantum} S.~Bugaichuk, A.~Kutana and  A.~Khiznyak, 
Quantum Electronics {\bf 27} 727 
(1997).

\bibitem{Sturman2005} B.~Sturman, E.~Povidilov and M.~Gorkunov, 
Phys.~Rev.~E {\bf 72} 016621 
(2005).  

\bibitem{BCPRE2009} S.~Bugaychuk and R.~Conte,
Phys.~Rev.~E {\bf 80} (2009) 066603. 

\bibitem{BKMPR} S.~Bugaychuk, L.~K\'ovacs, G.~Mandula, K.~Polg\'ar
 and R.A.~Rupp,                 
Phys.~Rev.~E {\bf 67}, 046603. 

\bibitem{JBHJOSAB} M.~Jeganathan, M.C.~Bashaw and L.~Hesselink,
\hfill\break\noindent
J.~Opt.~Soc.~Am.~B {\bf 12} 1370 
(1995).

\bibitem{CBJPA2009} R.~Conte and S.~Bugaychuk,
J.~Phys.~A: Math.~Theor.~{\bf 42} (2009) Fast Track Comm.~192003. 

\bibitem{CBScicli} R.~Conte and S.~Bugaychuk,
177, 
\hfill\break\noindent 
{\it Waves and stability in continuous media},
eds.~N.~Manganaro, R.~Monaco and S.~Rionero
(World scientific, Singapore, 2008).
http://arXiv.org/abs/0806.1183

\bibitem{DauxoisPeyrard} T.~Dauxois and M.~Peyrard, 
\textit{Physics of solitons},
422 pages (Cambridge University Press, Cambridge, 2006). 

\end{thebibliography}

\vfill\eject\end{document}